\documentclass[aps,pra,twocolumn,reprint,superscriptaddress,floatfix]{revtex4-2}

\usepackage{bm,bbm,amsmath,bbold,amssymb,amsbsy,txfonts,graphicx,color,hyperref,xfrac}
\usepackage{textgreek}
\usepackage{CJK}
\usepackage{todonotes}

\usepackage{float}
\usepackage{braket}
\usepackage{upgreek}

\usepackage[compact]{titlesec}           
\titlespacing{\section}{0pt}{20pt}{10pt} 

\usepackage[final]{pdfpages}

\makeatletter
\AtBeginDocument{\let\LS@rot\@undefined}
\makeatother

\usepackage{lineno}

\begin{document}

\title{Direct electrical modulation of surface response in a single plasmonic nanoresonator\\}


\author{Luka Zurak}
\email{luka.zurak@physik.uni-wuerzburg.de}
\affiliation{Nano-Optics and Biophotonics Group, Experimental Physics 5, Institute of Physics, University of Würzburg, Germany}
\author{Christian Wolff}
\affiliation{POLIMA -- Center for Polariton-driven Light-Matter Interactions, University of Southern Denmark, Campusvej 55, DK-5230 Odense M, Denmark}
\author{Jessica Meier}
\affiliation{Nano-Optics and Biophotonics Group, Experimental Physics 5, Institute of Physics, University of Würzburg, Germany}
\author{Ren\'e Kullock}
\affiliation{Nano-Optics and Biophotonics Group, Experimental Physics 5, Institute of Physics, University of Würzburg, Germany}  
\author{N. Asger Mortensen}
\affiliation{POLIMA -- Center for Polariton-driven Light-Matter Interactions, University of Southern Denmark, Campusvej 55, DK-5230 Odense M, Denmark}
\affiliation{Danish Institute for Advanced Study, University of Southern Denmark, Campusvej 55, DK-5230 Odense M, Denmark}
\author{Bert Hecht}
\affiliation{Nano-Optics and Biophotonics Group, Experimental Physics 5, Institute of Physics, University of Würzburg, Germany}
\author{Thorsten Feichtner}
\email{thorsten.feichtner@physik.uni-wuerzburg.de}
\affiliation{Nano-Optics and Biophotonics Group, Experimental Physics 5, Institute of Physics, University of Würzburg, Germany}

\maketitle

{\bf 

Classical electrodynamics describes the optical response of systems using bulk electronic properties and treats boundaries between two materials as infinitesimally thin. However, due to the quantum nature of electrons, interfaces have a finite thickness. Non-classical surface effects become increasingly important as ever smaller nanoscale systems are realized and eventually dominate over volume-related phenomena. Investigating the response of surface electrons in such systems, therefore, becomes imperative. One way to gain control over non-classical interface effects and study them is through electrical gating, as the static screening charges reside exclusively at the surface.
Here, we investigate the modulation of the surface response upon direct electric charging of a single plasmonic nanoresonator by measuring the resulting changes in resonance. We analyze the observed effects within the general framework of surface-response functions and provide a basic model derived from electron spill-out within the local-response approximation (LRA). Our observed change in resonance frequency is well accounted for by assuming a modulation of the in-plane surface current. Surprisingly, we also measure a change in the resonance width, where adding electrons to the surface leads to a narrowing of the plasmonic resonance, i.e., reduced losses. The description of such effects requires considering nonlocal effects and the inclusion of a possible anisotropy of the perturbed surface permittivity.
Our experiment, therefore, opens a vast field of investigations on how to gain control over the surface response in plasmonic resonators and to develop ultrafast and extremely small electrically driven plasmonic modulators and metasurfaces by leveraging electrical control over non-classical surface effects.}

\section{Introduction}
The interaction between free electrons at a metal surface and electromagnetic fields at optical frequencies gives rise to surface plasmon polaritons (SPPs), which are strongly coupled states of light and charge-density oscillations\,\cite{ritchie1957plasma}. When SPP waves are confined to finite-sized objects, they exhibit resonances with large field enhancements in the vicinity of maxima in the surface-charge density\,\cite{muhlschlegel2005resonant}. By tailoring the geometry of the object, one can effectively control and adjust the spectral position of the SPP resonances\,\cite{novotny2007effective}. These practical properties of localized plasmonic resonances have been extensively studied and widely utilized\,\cite{curto2010unidirectional,knight2011photodetection,zhang2013chemical,kern2015electrically,celebrano2015mode,chikkaraddy2016single}. However, the lack of possibilities to \emph{actively} tune plasmonic resonances still remains a major obstacle in realizing the full potential of plasmonic systems. Active tuning would enable the development of ultrafast and ultrasmall optical modulators and tunable metamaterials \,\cite{jiang2017active}. At the first glance, one of the most compelling pathways towards dynamic tuning is varying the density of the electron gas. This feature is already exploited in dilute systems, such as graphene plasmonics\,\cite{wang2008gate,bonaccorso2010graphene,liu2011graphene}, but often disregarded in metallic nanoresonators with higher carrier densities. To investigate the tuning of plasmonic nanoresonators various methods have been employed for enhancing the electrical charging, such as embedding resonators in a chemical reductant or an ion gel, which has resulted in a significant increase in the capacitance of the system\,\cite{mcmillan2007reflectance,mulvaney2006drastic,novo2009electrochemical,miyazaki2009electrical,dondapati2012voltage,petach2014mechanism,byers2014single,brown2015electrochemical,Mulvaney,hoener2017spectral,liu2018electrochemical,maniyara2019tunable}. However, these approaches have severe limitations, including slow operation and inconclusive results. Hysteresis has been observed in some cases, likely due to double-layer formation or electrochemical reactions at the metal surface, suggesting different origins of the observed tuning effects beyond pure charging. Consequently, there is a necessity to carry out experiments free from the potential impacts of environmental variations by means of direct electrical charging.

On the theoretical side a common approach in modeling of observed resonance changes is based on the incorrect assumption that electrostatic charges lead to a change in the bulk properties\,\cite{mulvaney2006drastic,novo2009electrochemical,sheldon2014plasmoelectric}. Only few studies have explored the impact of charging on the optical response, viewing it as a surface-related phenomenon \,\cite{bohren_scattering_1977,brown2015electrochemical,hoener2017spectral,zapata2016plasmon,li2022direct}. Among these, Brown \emph{et al.} utilize a core-shell model, offering only a simplistic explanation and, consequently, limited insights into the optical characteristics of electrons at the surface\,\cite{brown2015electrochemical}. While Hoener \emph{et al.}~did develop a more advanced semi-classical model based on electron spill-out, it ultimately resulted in a 
debated 
interpretation of the experimental observations \cite{hoener2017spectral}. 
In the latest developments, Li \emph{et al.} have examined the resonance of a nanoscale electron reservoir placed within a narrow plasmonic gap using a quantum hydrodynamic model\,\cite{li2022direct}. Their theoretical findings show enormous potential, emphasizing the possibility to obtain large resonance modulation driven by a strong perturbation of the out-of-plane surface electron response, while neglecting the in-plane response. We conclude that a universal framework and a simple analytical model, to treat changes of the optical response of plasmonic nanoresonators upon charging, is unavailable. In particular, taking into account the full non-classical properties of electrons at the surface, remains a challenge. It is no coincidence that more recently, the further development of a framework to describe the optical response of electrons at the surface of metals has gained much attention, as non-classical surface properties have been found to strongly shape the response of nanoscale plasmonic systems\,\cite{feibelman1982surface,kreibig1985optical,tiggesbaumker1993blue,cottancin2006optical,zuloaga2009quantum,savage2012revealing,scholl2012quantum,savage2012revealing,ciraci2012probing,raza2013blueshift,zhu2016quantum,yang2019general,boroviks2022}. Reduced field enhancement, accompanied by broadening and spectral shifting of resonances, are caused by electronic spill-out\,\cite{zhu2016quantum,mortensen2021surface}, nonlocality\,\cite{raza2015nonlocal}, and Landau damping\,\cite{jin2015quantum,khurgin2015ultimate,shahbazyan2016landau}, effects which can be treated using the universal framework of surface-response functions, better known as Feibelman $d$-parameters\,\cite{yang2019general}. 

As nanofabrication and computational methods continue to progress, the potential and importance of smaller nanoresonators with pronounced surface effects are increasing, bridging the divide between classical and quantum facets of linear and nonlinear optical responses. Therefore, conducting experiments that employ direct electrical charging and evaluating the results within the framework of surface-response functions can contribute to a deeper understanding, ultimately facilitating the enhancement of electrical tuning in smaller systems where surface effects play a dominant role. Moreover, the appearance of an inverse optoelectronic effect, referred to as the plasmoelectric potential, hinges fundamentally on the electrical tunability of plasmonic resonances \,\cite{sheldon2014plasmoelectric}. 

\indent In this study, we experimentally investigate the effect of direct electrical charging on the fundamental dipolar resonance of a single plasmonic nanoresonator (as sketched in Fig.~\ref{fig:fig1}a), while carefully excluding slow phenomena typically associated with electrochemical processes. To this end, we measure the relative change in light scattering for applied electric potentials of up to $\pm$20\,V, from which we determine the induced changes in the nanoresonantor's eigenfrequency and line width. We place the observed changes within the general framework of surface-response functions for which we provide a basic model derived from electron spill-out in local-response approximation (LRA). 
The observed shift in the resonance frequency can be nicely explained by a modulation of the in-plane surface response. However, we also observe a change in the resonance width, which is in stark contrast with the predictions of our model. To account for this observation the basic assumptions of our model need to be extended likely towards anisotropy of the local permittivity, the perturbation of conductivity via surface-states, or nonlocal effects hidden in the perpendicular component of the $d$-parameters.

\section{Surface-response functions}
\begin{figure}[h!]
	    \centering	    
            \includegraphics[width=.5\textwidth]{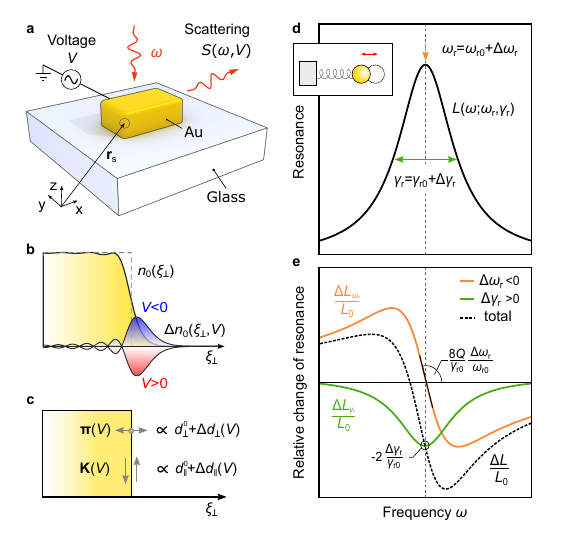}
            \caption{\textbf{Influence of electrostatically induced surface charges on optical response of plasmonic nanoresonators.} \textbf{a}, Sketch of the system under research; a rectangular gold (Au) nanoresonator is placed on top of a glass substrate. The applied electric potential $V$ induces a change in the resonance which is detected by recording the scattering signal $S(\omega,V)$ as a function of the photon energy $\hbar\omega$ and the applied potential. \textbf{b}, Equilibrium and induced electron density distributions at the gold-air interface calculated using a Wigner--Seitz radius of $r_\text{s}/a_0=3.18$ in the jellium description. Grey-dashed line denotes the position of a positive background. \textbf{c}, The change in electron density at the surface position $\textbf{r}_\text{s}$ perturbs the $d$-parameters, and in turn the optical response of the system by altering the surface polarization $\textbf{P}_\text{s} (V)$ (see Eq.~\eqref{eq:eigenfrequency_perturbation}). \textbf{d}, Model of a harmonic oscillator and a corresponding resonance curve with a Lorentzian line shape $L(\omega;\omega_\mathrm{r},\gamma_\mathrm{r})$ described by its resonant frequency $\omega_\mathrm{r}$ and width $\gamma_\mathrm{r}$. \textbf{e}, Small perturbations of $\Delta\omega_\mathrm{r}<0$, or $\Delta\gamma_\mathrm{r}>0$, lead to distinct relative changes of the resonance represented by the orange curve ($\Delta \omega_\mathrm{r}$) and the green curve ($\Delta \gamma_\mathrm{r}$), respectively. Change in resonance width is related to relative change at the resonant frequency, while change in resonance position is linked to the slope of this relative change - $Q=\omega_\mathrm{r}/\gamma_\mathrm{r}$ stands for the quality factor. Total change, represented by the black dashed line, is a superposition of the two.}
	    \label{fig:fig1}
\end{figure}

Bohren and Hunt\,\cite{bohren_scattering_1977} made one of the initial attempts to model how adding electrons to the surface of a plasmonic nanoresonator affects the resonance. They assumed that, in the picture of classical electrodynamics, added electrons modify the surface conductivity $\sigma_\text{s}$ and consequently perturb the in-plane surface current $\textbf{K}=\sigma_\text{s}  \textbf{E}_\parallel$ driven by the tangential component of electric field $\textbf{E}_\parallel$ yielding a change in the effective plasma frequency (see Supplementary Section 1.1). However, to fully capture the optical properties of electrons at the surface one has to take into account various non-classical effects. First of all, electron spill-out causes the surface to acquire a finite thickness (see Fig.~\ref{fig:fig1}b) and leads to an out-of-plane response. Other quantum effects which influence the electronic response at the surface are nonlocality, Landau damping and conduction through surface-states\,\cite{mortensen2021mesoscopic,echarri2021optical}. All such effects can be accounted for by Feibelman $d$-parameters\,\cite{feibelman1982surface,mortensen2021mesoscopic}. These $d$-parameters, $d_{\perp}$ and $d_{\parallel}$, are centroids of the induced charge density and of the normal derivative of the tangential current, respectively \cite{feibelman1982surface} (see Supplementary Section 1.2-3). Hence, as demonstrated by Christensen \emph{et al.}\,\cite{christensen2017quantum}, $d$-parameters describe the surface polarization
\begin{equation}
\centering
\label{eq:surface_polarization}
\textbf{P}_\text{s}=\boldsymbol{\uppi}+i\omega^{-1}\textbf{K},
\end{equation}
which can be added to the classical framework by extending the boundary conditions \cite{yang2019general}. The out-of-plane electron oscillation, proportional to $d_{\perp}$, is described via the dipole density $\boldsymbol{\uppi}=\varepsilon_0 d_\perp \Delta E_\perp \hat{x}_\perp$, where $\varepsilon_0$ is the free-space permittivity, $E_\perp$ is the perpendicular component of electric field, and $\hat{x}_\perp$ is the unit vector normal to the surface. The parallel component $d_{\parallel}$ contributes to an in-plane surface current $\textbf{K}=i\omega d_\parallel \Delta \textbf{D}_\parallel$ as already discussed above, where $\textbf{D}_\parallel$ is the tangential component of the displacement field (see Fig.~\ref{fig:fig1}c). 

If we assume a simple harmonic oscillator as a model for a plasmonic nanoresonator (see Fig.~\ref{fig:fig1}d), the associated Lorentzian $L(\omega;\omega_\mathrm{r},\gamma_\mathrm{r})$ can be described by a complex eigenfrequency $\Tilde{\omega}_\mathrm{r}=\omega_\mathrm{r}-i\gamma_\mathrm{r}/2$ \cite{lalanne2018light}. The real part corresponds to the resonance frequency, while the imaginary part describes the attenuation of the system. It contains radiation as well as Ohmic losses. Surface effects described by the Feibelman $d$-parameters lead to a change of the plasmonic nanoresonator's complex eigenfrequency $\Delta \Tilde{\omega}_\mathrm{r}$, which is directly related to the surface polarization via the integral\,\cite{yang2015simple}
\begin{equation}
\centering
\label{eq:eigenfrequency_perturbation}
\Delta\Tilde{\omega}_\mathrm{r}=-\Tilde{\omega}_\text{r0}\int_\mathrm{s}\textbf{E}^{(0)}\cdot \textbf{P}^{(0)}_\text{s}\,\mathrm{d}s,
\end{equation}
where $\mathrm{s}$ denotes the surface of the resonator. Here, $\textbf{P}_\text{s}^{(0)}$ and $\textbf{E}^{(0)}$ are evaluated at the unperturbed complex eigenfrequency $\Tilde{\omega}_\text{r0}$ obtained by assuming classical piecewise constant material properties, neglecting any non-classical effects introduced via the surface polarization $\textbf{P}^{(0)}_\text{s}$. Therefore, the experimentally observed eigenfrequency of an uncharged plasmonic nanoresonator corrected by surface effects is given as a $\Tilde{\omega}_\mathrm{r}=\Tilde{\omega}_\text{r0}+\Delta\Tilde{\omega}_\mathrm{r}$. 

In the case of a controlled perturbation, the additional shift in the spectral position of the resonance frequency $\Delta\omega_\mathrm{r}$ produces a distinct pattern in the relative change of the resonance, as depicted in Fig.~\ref{fig:fig1}e. It can be easily differentiated from the relative change pattern caused by a perturbation of the width of the resonance $\Delta\gamma_\mathrm{r}$ (see Supplementary Section\,1.4). 

We use this formalism to investigate and describe the charge-induced eigenfrequency shifts of a single gold nanoresonator residing on top of a glass substrate (see Fig.~\ref{fig:fig1}a). At any point on the surface of the nanoresonator $\textbf{r}_\text{s}$, an applied voltage $V$ will introduce a change in equilibrium electron density (see Fig.~\ref{fig:fig1}b) and locally perturb the surface-response functions  
\begin{align}
    \label{eq:eq3}
    \begin{split}
    d_{\perp}(\textbf{r}_\text{s},V)\simeq d^{0}_{\perp}(\textbf{r}_\text{s})+&\frac{\partial d_{\perp}(\textbf{r}_\text{s},V)}{\partial V} \; V,\\
    d_{\parallel}(\textbf{r}_\text{s},V) \simeq d^{0}_{\parallel}(\textbf{r}_\text{s})+ & \frac{\partial d_{\parallel}(\textbf{r}_\text{s},V)}{\partial V} \; V.
\end{split}
\end{align}
Here, $d^{0}_{\perp}(\textbf{r}_\text{s}), d^{0}_{\parallel}(\textbf{r}_\text{s})$ are unperturbed $d$-parameters, typically taken to be constant across a metal-dielectric interface. The $d$-parameter perturbations $\Delta d_{\perp}(V), \Delta d_{\parallel}(V)$ will lead to a local change of the surface polarization $\Delta\textbf{P}_\text{s}(\textbf{r}_\text{s},V)$ (see Fig.~\ref{fig:fig1}c), which will introduce a voltage-induced change in the system's complex eigenfrequency $\Delta \Tilde{\omega}_\mathrm{r}(V)$ in accordance with Eq.~\eqref{eq:eigenfrequency_perturbation}. These changes 
can be detected by recording the relative change of scattering $\Delta S (V)/S_{\!0}$ (as shown in Fig.~\ref{fig:fig1}e), where $\Delta S(V)$ is the voltage-induced change in scattering signal and $S_{\!0}$ is the scattering signal of the uncharged system -- the scattering signal is defined as normalized scattering cross section $S\propto\sigma_\mathrm{sca}$.
 
\section{Experiment}
\begin{figure*}[t]
	\centering
	    \includegraphics[width=.99\linewidth]{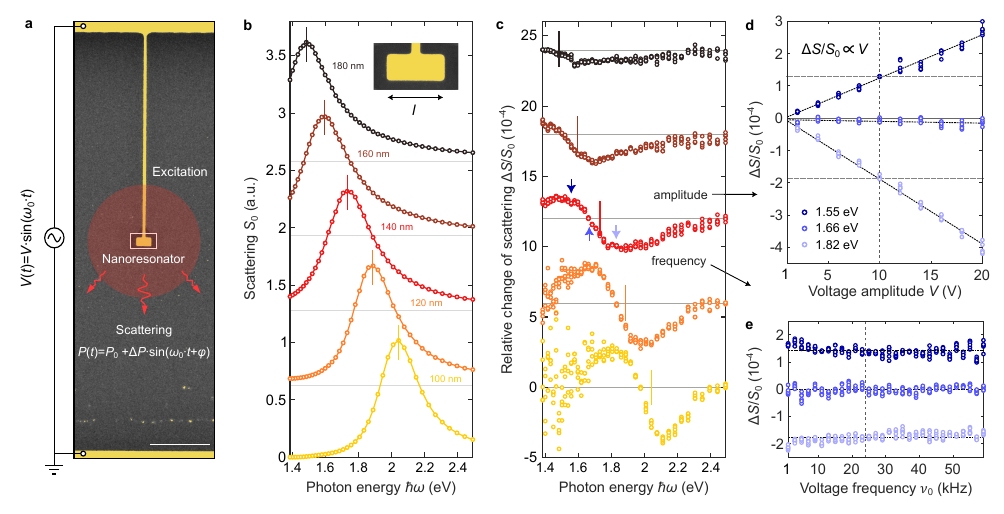}
	    \caption{\textbf{Measuring the voltage induced change of a plasmonic resonance.} \textbf{a}, Colored SEM image of a single electrically connected gold nanoresonator. The structure is driven with a sinusoidal voltage $V(t)$ while monitoring the scattered power $P(t)$. Scale bar is 500\,nm. \textbf{b}, Measured scattering spectra for resonators with length ranging from 180\,nm down to 100\,nm - due to the small error margin, which hinders the discrimination of individual points, only one point per photon energy is plotted. The inset shows a colored close-up SEM image of the 180\,nm long nanoresonator. \textbf{c}, Relative change of scattering as obtained from lock-in measurements for 10\,V of voltage amplitude corresponding to the resonators in \textbf{a}. \textbf{d}(\textbf{e}), Dependency of relative change of scattering on the amplitude (frequency) of the applied voltage, respectively, for three spectral points denoted in \textbf{c}.}
	    \label{fig:fig2}
\end{figure*}
To experimentally investigate the effect of charging on the plasmonic resonance, we conducted dark-field scattering measurements on electrically connected nanoresonators, as depicted in Fig.~\ref{fig:fig2}a. The plasmonic resonators under investigation are fabricated from 50\,nm thick single crystalline gold microplatelets\,\cite{krauss2018controlled}, using a two-step focused ion beam (FIB) milling process (similar to the procedure in Ref.\,\onlinecite{meier2022controlling}). The first step involved Ga-FIB to create the rough shape of the resonator and connector. Then, He-FIB was employed to refine the structure's final shape with precision. An example of a 180\,nm long nanoresonator is shown in the inset of Fig.~\ref{fig:fig2}b. The shape of the resonance was probed with a supercontinuum white-light laser, spectrally tuned using a narrowband filter, with its collimated beam polarized along the long-axis of the resonator (for more details see Supplementary Section\,2.1). The scattering signal $S$ is proportional to the ratio of the power scattered by the structure $P$ and the incoming intensity $I$ (see Supplementary Section\,2.2-3). In Fig.~\ref{fig:fig2}b, we plot the normalized scattering spectra of 80\,nm wide resonators with lengths ranging from 180\,nm down to 100\,nm. 
To retrieve the change in scattering 
we drive the system using a sinusoidal voltage with an amplitude $V$ at frequency $\nu_\mathrm{0}$ and employ a phase-sensitive lock-in amplifier to detect both the amplitude $\Delta P$ and the phase $\phi$ of the voltage-modulated scattered power, as illustrated in Fig.~\ref{fig:fig2}a. The unperturbed scattered power $P_0$ was recorded using an optical powermeter placed in one of the branches of the detection system (see Supplementary Figure\,S4). 
By repeating the measurement across the spectrum for a voltage amplitude of 10\,V at a linear frequency of 24\,kHz, we observe a purely in-phase lock-in signal ($\phi$ = {0, $\pi$}) at the fundamental frequency for the relative change of scattering as shown in Fig.~\ref{fig:fig2}c (see Supplementary Section\,2.4). 

\section{Results and discussion}
Across all nanoresonators, we consistently observe a similar spectral shape in the change of scattering, suggesting a red-shift and broadening of the resonance under positive bias, i.e. exhibiting a blue-shift and narrowing under negative bias. In contrast to classical models which predict solely a shift of the resonance frequency we observe also an influence on the resonance width, where adding electrons to the surface reduces the loss. Notably, smaller structures demonstrate a more pronounced modulation indicating a route to further increase the effect. 
For the exemplary case of the 140\,nm long resonator, we demonstrate a linear relation between the observed changes and the applied voltage's amplitude by recording signals at three specific spectral points (see Fig.~\ref{fig:fig2}d). Additionally, we demonstrate that the observed changes at these three spectral points remain unaffected by the frequency of the AC signal within the observable spectral range, constrained by the responsiveness of the lock-in amplifier (see Fig.~\ref{fig:fig2}e). This implies that we can exclude slow processes typically encountered in the experiments performed in electrochemical cells. 
To get an insight into the mechanism which causes the experimentally observed resonance changes within the framework of surface-response functions, we need to determine the voltage-dependent $d$-parameter perturbations (see Eq.~\eqref{eq:eq3}).

\subsection{Electron spill-out and local response approximation (LRA)}
\begin{figure*}[]
\centering
\includegraphics[width=.99\textwidth]{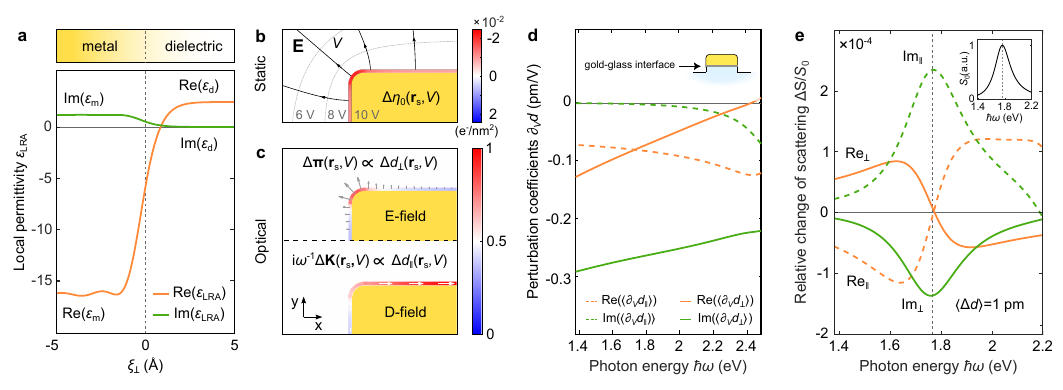}
  \caption{\textbf{Influence of $d$-parameter perturbations on a plasmonic resonance.} \textbf{a}, Spatial dependency of local permittivity at the gold-glass interface. \textbf{b}, Simulated static electric field lines at half of the antenna height (black arrowed lines) and contours of electric potential (gray dashed lines) for a 10\,V equipotential at the surface of the structure. Excess electrons are non-uniformly distributed across the surface, represented by the induced surface electron density $\Delta\eta_\text{0}(\textbf{r}_\text{s},V)=C_\mathrm{s}(\textbf{r}_\text{s})V/q_\mathrm{e}$ (surface color). \textbf{c}, Simulated perturbed surface polarization $\textbf{P}_\text{s}(V)$ corresponding to an out-of-plane response (up) with electric field, and an in-plane response(down) with displacement field. \textbf{d}, Spectral dependency of surface averaged $d$-parameter perturbation coefficients for the gold-glass interface calculated with Eq.~\eqref{eq:dpara} and \eqref{eq:dperp}. The $d_\perp$ component contributions are depicted with a solid lines, while the $d_\parallel$ component are represented with dashed lines. Real parts are given in orange, and imaginary parts in green. \textbf{e}, Calculated relative change of scattering $\Delta S/S_{\!0}$ for 1\,pm of introduced surface-averaged $d$-parameter perturbations. The inset shows the calculated unperturbed scattering spectrum $S_{\!0}$.}
  \label{fig:fig3}
\end{figure*}

A complete analysis in determining $d$-parameter perturbations would encompass all possible microscopic phenomena taking place at the surface of the charged nanoresonator. Here, in a first approximation, we calculate the perturbation coefficients $\partial_V d_\perp$ and $\partial_V d_\parallel$, as a function of the position on the  surface $\textbf{r}_\text{s}$, resulting solely from the electron spill-out under the LRA. We assume a position-dependent plasma frequency $\omega_\text{p}(\xi_\perp,V)=\omega_\text{p0}\sqrt{n_\text{0}(\xi_\perp,V)/n_\text{0}}$, where $\omega_\text{p0}$ is the unperturbed plasma frequency, $n_\text{0}(\xi_\perp,V)$ is voltage-dependent equilibrium electron density, $n_0$ is the bulk density of free electrons and $\xi_\perp$ parameterizes the distance perpendicular to the interface to capture e.g.~the electron spill-out. In this approximation, the optical properties are then described by a position-dependent local permittivity (see Fig.~\ref{fig:fig3}a) \begin{equation}
\label{eq:local_permittivity}
\varepsilon_\mathrm{LRA}(\xi_\perp,V)=\varepsilon_\mathrm{b}(\xi_\perp)-\frac{\omega_\mathrm{p0}^2}{\omega^2+i\gamma\omega}\frac{n_\text{0}(\xi_\perp)+\Delta n_\text{0}(\xi_\perp,V)}{n_\mathrm{0}},
\end{equation}
where $n_\text{0}(\xi_\perp)$ is the equilibrium electron density of an uncharged system and $\Delta n_\text{0}(\xi_\perp,V)$ is the voltage-induced electron density (see Fig.~\ref{fig:fig1}d). $\varepsilon_\mathrm{b}(\xi_\perp)$ is the position dependent background permittivity which in bulk gold is dominated by interband contribution while outside the gold it transitions to the permittivity of the surrounding dielectric. For simplicity, we assume that the electron response at any point at the surface is isotropic, although it is expected that the equation-of-motion perpendicular to the interface contains additional terms that describe the influence of the boundary. Within these assumptions, according to Ref.\,\onlinecite{mortensen2021surface}, the $d$-parameter perturbation coefficients can be expressed in integral form as
\begin{subequations}
\begin{align}
    \frac{\partial d_{\perp}}{\partial V}=&\frac{\varepsilon_\text{d}\varepsilon_\text{m}}{\varepsilon_\text{m}-\varepsilon_\text{d}}\int_{-\infty}^{\infty}\displaylimits \mathrm{d}\xi_\perp \frac{1}{\varepsilon_\text{LRA}^2(\xi_\perp,V)}\frac{\partial}{\partial V}\varepsilon_\text{LRA}(\xi_\perp,V), \label{eq:eperturbation_coefficient_perpendicular} \\ 
    \frac{\partial d_{\parallel}}{\partial V}=&\frac{1}{\varepsilon_\text{m}-\varepsilon_\text{d}}\int_{-\infty}^{\infty}\displaylimits \mathrm{d}\xi_\perp \frac{\partial}{\partial V}\varepsilon_\text{LRA}(\xi_\perp,V). \label{eq:eperturbation_coefficient_parallel}
\end{align}
\end{subequations}
Here, $\varepsilon_\text{m}$ and $\varepsilon_\text{d}$ are the bulk permittivities of the metal and surrounding dielectric, respectively (for a detailed derivation see Supplementary Section\,1.5). 
For both components, the integrands in equations~\eqref{eq:eperturbation_coefficient_perpendicular} and\,\eqref{eq:eperturbation_coefficient_parallel} contain the partial derivative of the local permittivity with respect to the applied voltage $\partial_\mathrm{V}\varepsilon_\mathrm{LRA}$. That is, they depend on the induced electron density $\Delta n_0 (\xi_\perp,V)$, which can be related to the classical total induced surface electron density $\Delta \eta_0(V)$ via the integral 
\begin{equation}
\centering
\label{eq:induced_electron_density1}
\int_{-\infty}^{\infty}\displaylimits \mathrm{d}\xi_{\perp} \Delta n_0 (\xi_\perp,V)=\Delta \eta_0(V).
\end{equation}
The voltage-induced surface electron density can be obtained from the discontinuity of the normal component of the static displacement field $D_\perp=\varepsilon_\mathrm{0} \kappa E_\perp$, where $E_\perp$ is the normal component of the static electric field (see Fig.~\ref{fig:fig3}b) in the dielectric and $\kappa$ is the dielectric constant. Furthermore, we can express the induced surface electron density in terms of a surface capacitance $C_\text{s}$ as $\Delta\eta_\text{0}(V)\simeq C_\text{s} V /q_\text{e}$, where $q_\text{e}$ is the electron charge, showing explicitly the dependency on the applied voltage. This implies that we can represent the induced electron density as 
\begin{equation}
\label{eq:induced_electron_density2}
\Delta n_0 (\xi_\perp,V)\simeq p(\xi_\perp)\frac{C_\mathrm{s} V}{q_\mathrm{e}}, 
\end{equation}
where $p(\xi_\perp)$ is the probability distribution of the induced electron density per unit length, which follows the shape of the induced electron density (see Fig.~\ref{fig:fig1}e) and satisfies the normalization condition $\int\mathrm{d}x_\mathrm{\perp}p(\xi_\perp)=1$. In the case of the $d_\parallel$ component, using equation~\eqref{eq:local_permittivity} for the evaluation of equation~\eqref{eq:eperturbation_coefficient_parallel} reduces to the evaluation of the partial derivative with respect to voltage of the integral  Eq.~\eqref{eq:induced_electron_density1}. This expression can be evaluated by employing equation~\eqref{eq:induced_electron_density2}, which implies that the perturbation is independent of the shape of the induced electron density and only depends on the amount of the induced charge. Therefore, the perturbation coefficient can be expressed analytically with a purely classical term
\begin{align}
\label{eq:dpara}
\begin{split}
\frac{\partial d_{\parallel}(\textbf{r}_\text{s},V)}{\partial V}\simeq&\frac{1}{\varepsilon_\text{d}-\varepsilon_\text{m}}\frac{\omega_\text{p}^2}{\omega^2+i\gamma\omega}\frac{C_\text{s}(\textbf{r}_\text{s})}{q_\text{e}}\frac{1}{n_\text{0}},
\end{split}
\end{align}
where the surface position dependency $\mathbf{r}_\mathrm{s}$ has been reintroduced through the surface capacitance $C_\text{s}(\textbf{r}_\text{s})$, arising from the uneven distribution of static charges. One can show that the perturbed $d_\parallel$ parameter directly leads to a perturbed surface conductivity $\Delta \sigma_\text{s}\propto \Delta d_\parallel$, the same result as provided by the classical model of Bohren and Hunt (for a detailed discussion see Supplementary Section\,1.6). In addition to the perturbation of the surface conductivity resulting from the bulk electrons, other, non-classical effects, such as conduction through surface-states, can also impact the $d_{\parallel}$ perturbation coefficient\,\cite{mortensen2021mesoscopic,echarri2021optical}, especially since our resonators are fabricated from single crystalline gold microplatelets with Au(111) top and bottom surfaces \cite{yan2015topological,dreher2023focused}. 

In the case of the $d_\perp$ component, by inserting equation~\eqref{eq:induced_electron_density2} for the induced electron density into equation~\eqref{eq:local_permittivity}, the surface position dependency carried by the surface capacitance can be brought outside the integral in equation~\eqref{eq:eperturbation_coefficient_perpendicular} which results in the following expression
\begin{align}
\begin{split}
\centering
\label{eq:dperp}
    \frac{\partial d_{\perp} (\textbf{r}_\mathrm{s}, V)}{\partial V} \approx \frac{\varepsilon_\mathrm{d}\varepsilon_\mathrm{m}}{\varepsilon_\mathrm{d}-\varepsilon_\mathrm{m}} \frac{\omega_\mathrm{p}^2}{\omega^2+i\gamma\omega}\frac{C_\mathrm{s}(\textbf{r}_\mathrm{s})}{q_\mathrm{e}}\frac{1}{n_{0}} \int_{-\infty}^{\infty}\displaylimits d\xi_\perp \frac{p(\xi_\perp)}{\varepsilon_\mathrm{LRA}^2(\xi_\perp)}.
\end{split}
\end{align}
Evaluation of the integral~\eqref{eq:dperp} is not trivial since it depends on the shape of the induced electron density and the local permittivity. For simplicity, we assume that the background permittivity \eqref{eq:local_permittivity} follows the distribution of the equilibrium electron density (see Fig.~\ref{fig:fig3}a), which is derived from DFT calculations using the jellium approximation. Subsequently, as one transitions from the metal to the dielectric, the equilibrium electron density smoothly changes from the bulk value to zero. This results in a position, $\xi_{\perp,o} $, where the local permittivity (real part) crosses zero. Therefore, accurately evaluating the integral in equation~\eqref{eq:dperp} necessitates a precise knowledge of the shape of the permittivity in the vicinity of $\xi_{\perp,o}$.

\subsection{Simulations}
\indent To evaluate the perturbation coefficients ~\eqref{eq:dpara} and \eqref{eq:dperp}, we need to evaluate the equilibrium $n_\text{0}(\xi_\perp)$, and the induced electron density $\Delta n_\text{0}(\xi_\perp,V)$, as well as the position dependent surface capacitance $C_\text{s}(\textbf{r}_\text{s})$. To this end we perform DFT calculations for a thin slab of jellium (8\,nm), and electrostatic simulations for the geometry under investigation (for more details refer to Supplementary Section\,1.7-8). In Fig.~\ref{fig:fig3}d we show the spectrally dependent, interface averaged $d$-parameter perturbation coefficients for 140\,nm long resonator. Real and imaginary parts for all the perturbation coefficients exhibit negative values and show gradual variations across the spectrum. These coefficients have magnitudes on the order of 0.1\,pm/V. Therefore, for typical voltages of around 10\,V, the resulting perturbations $\braket{\Delta d}=\braket{\partial_V d} V$ will be approximately 1\,pm. Negative signs in the real parts are expected, since for a positive bias we deplete the surface from polarizable material, effectively reducing the size of the resonator. Although in Fig.~\ref{fig:fig3}d we show only the case of the gold-glass interface, the result is similar for the gold-air interface. 

Next, we examine the extent to which each component of the $d$-parameters influences the scattering spectrum. First, numerical simulations are conducted to obtain the scattering resonance as shown in the inset of Fig.~\ref{fig:fig3}e. In the second step, we vary each of the $d$-parameters independently by $1$\,pm of surface-averaged value and calculate the relative change of scattering $\Delta S/S_{\!0}$ using mesoscopic boundary conditions (see Fig.~\ref{fig:fig3}e)\,\cite{yan2015PRL,christensen2017quantum,yang2019general}. We find that the calculated changes are of the same order of magnitude as the experimentally observed ones. Furthermore, we can see that for a real part perturbation of 1\,pm in the $d_\parallel$($d_\perp$) component (denoted as $\mathrm{Re}_\parallel$($\mathrm{Re}_\perp)$ in Fig.~\ref{fig:fig2}c, respectively) we obtain dispersive curves with maximal changes on the slopes of the resonance, and a zero-crossing point almost perfectly aligned with the position of the resonance frequency. These characteristic shapes of the relative changes are produced by a change in the resonance position (compare orange curves in Fig.~\ref{fig:fig3}d and Fig.~\ref{fig:fig1}e). Moreover, introducing a perturbation of 1\,pm to the imaginary part of the $d_\parallel$($d_\perp$) component (denoted as $\mathrm{Im}_\parallel$($\mathrm{Im}_\perp$) in Fig.~\ref{fig:fig2}c) predominantly influences the width of the resonance (compare green curves in Fig.~\ref{fig:fig3}c and Fig.~\ref{fig:fig1}e). We can see that the induced changes in scattering for two parameters are of similar magnitude but act in the opposite directions while the perturbation coefficients have the same sign, implying a competing effects. This can explain why, in the quantum approach by Li and co-workers, with a strong out-of-plane response, it is expected to observe a strong red-shift and resonance broadening for a negatively charged resonator\,\cite{li2022direct}. In contrast, the classical model proposed by Bohren and Hunt, solely influenced by an in-plane response, predicts a blue-shift.

\subsection{Comparison of experiment and model} 
\begin{figure}[]
	    \centering \includegraphics[width=0.99\linewidth]{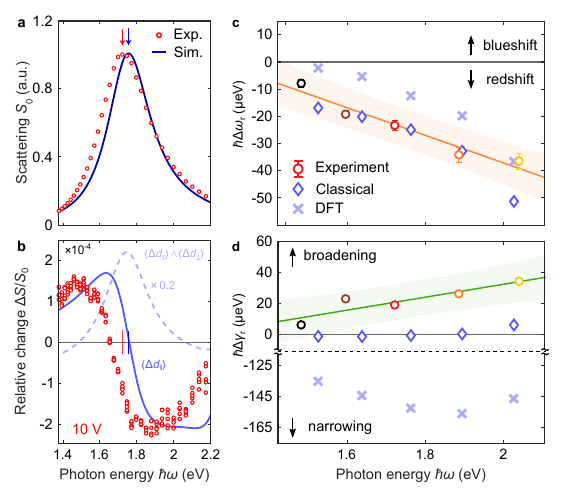}
	    \caption{\textbf{Comparison of experiment and model.} \textbf{a}, Scattering spectrum for 140 nm long resonator in experiment (red circles) 
     and simulation (dark blue line). 
     \textbf{b}, Relative change of scattering corresponding to resonances presented in \textbf{a} for a 10\,V of  voltage amplitude. Solid blue line corresponds to classical model while light-blue dashed line to model based on DFT. Small dashes represent the position of resonance frequency presented in \textbf{a}. \textbf{d,e}, Perturbations of resonant frequencies and resonance widths retrieved from experiments and simulations for both classical (blue diamonds) and DFT model (light blue crosses). Color code of the nanoresonators is identical to Fig.~\ref{fig:fig2}.}
	   \label{fig:fig4}
\end{figure}

By summing all $d$-parameter contributions to the relative change of scattering, we can determine the total change in resonance and directly compare it to experimental data, as presented in Fig.~\ref{fig:fig4}. To ensure that simulations and experiments are treated on an equal footing, the unperturbed resonance position $\omega_\mathrm{r0}$ and width $\gamma_\mathrm{r0}$ are obtained by fitting a Lorentzian line shape with a linear background to the scattering spectrum $S_\mathrm{0}$ (see Supplementary Section 1.9). Fig.~\ref{fig:fig4}a shows that the simulated curve closely matches the shape of the measured resonance. Minor discrepancies in the peak position and resonance width can be attributed to inaccuracies in the model's geometry and the material data. In Fig.~\ref{fig:fig4}b, alongside the experimental data, we show the relative change of scattering obtained from simulations for 10\,V of applied potential. We can see that the perturbation caused solely by the $d_\parallel$ component, which is, as pointed out, a purely classical term, exhibits a similar spectral shape as the experimental result. 
This similarity is reflected in the obtained spectral shift $\hbar\Delta\omega_\mathrm{r}$, which is determined to be $(-24.90\pm 0.01)\,\upmu$eV in the simulations, whereas the measured value is $(-23\pm 2)\,\upmu$eV -- we have used one standard deviation to represent the uncertainties. There is, however, a significant difference to observe when examining the point at which the relative change in scattering crosses zero. In the experimental data, this spectral position is shifted towards the blue compared to the resonance position, which suggests that the resonance has broadened (see Fig.~\ref{fig:fig1}c), with an increase in spectral width $\hbar\Delta\gamma_\mathrm{r}=(19\pm 1)\,\upmu$eV. This feature is not accounted for by the classical model -- the $d_\parallel$ component perturbation has negligible imaginary part (see Fig.~\ref{fig:fig2}d), and therefore does not affect the loss of the system, as reflected in the obtained $\hbar\Delta\gamma_\mathrm{r}$ value of $(0.7\pm 0.1)\,\upmu$eV. In contrast, when we include the $d_\perp$ component perturbation obtained from the DFT calculations, which is dominated by the large imaginary part (see Fig.~\ref{fig:fig3}d), we observe a maximum relative change of approximately $1 \times 10^{-3}$ at resonance. This implies a significant narrowing of the resonance for positive bias with a $\hbar\Delta\gamma_\mathrm{r}$ value of $(-152.8 \pm 0.1)\,\upmu$eV and reduced spectral shift $\hbar\Delta\omega_\mathrm{r}=(-12.4 \pm 0.4)\,\upmu$eV. However, none of the $d_\perp$-component effects have been observed in the experiment. Furthermore, as shown in Fig.~\ref{fig:fig4}c,d, both the resonances red-shifting and broadening occur for the resonators of different lengths, with scattering spectra presented in Fig.~\ref{fig:fig2}b. We can see that the classical model predicts the spectral shift well, however cannot account for the spectral broadening. The non-classical model based on DFT, which includes both in-plane and out-of-plane responses, cannot account for either.

\subsection{Discussion on the relative contributions}
The absence of the $d_\perp$-component effects, as calculated from DFT, suggests that contributions from the $d_\perp$ component are effectively suppressed in the experiment. Assuming that the DFT jellium calculations -- at least qualitatively -- captures the surface electrodynamic phenomena of the real physical surface, this leads to the conclusion that either nonlocal effects counteract the perturbation as calculated with the LRA, or there is an anisotropy of the perturbed local permittivity with suppressed out-of-plane response. If the out-of-plane motion is suppressed, one possible explanation for the narrowing of the resonance with added electrons is that added electrons preferentially occupy surface states and impact the $d_{\parallel}$ perturbation coefficient as our resonators are fabricated from single crystalline gold microplatelets with Au(111) top and bottom surfaces. 
To further validate these assumptions, additional experiments are required that can clearly separate the contributions of the nanoresonator's eigenfrequency change stemming from the perturbation of the $d_\parallel$ and $d_\perp$ component. Fig.~\ref{fig:fig3}e shows that the perturbation strengths for all components are non-negligible, which can be attributed to the simplicity of the chosen geometry that does not have any pronounced geometrical features and large field enhancements. One possible way to disentangle the contributions stemming from $d_\parallel$ and $d_\perp$, is by using a dimer antenna with a narrow gap\,\cite{prangsma2012electrically}. Such a structure exhibits a large field enhancement in the gap in both the static and optical regimes, which according to equation~\eqref{eq:eigenfrequency_perturbation}, should enhance contributions from the $d_\perp$ component. Nevertheless, due to anti-symmetric static perturbation, one should choose antenna with an asymmetric gap\,\cite{li2022direct,meier2022controlling}.

\section{Conclusion}
We have presented both theoretical and experimental investigation of charged plasmonic resonators and their optical response. In our experimental work, we have shown that direct electrical charging leads to a resonance shift and a change in the line width. Under the influence of a positive bias, the resonance shifts towards longer wavelengths while simultaneously increasing in width while the opposite effects occur under a negative bias. To explain the experimental observations, we place the charging effects within the general framework of surface-response functions taking into account both in-plane and out-of-plane responses. Feibelman $d$-parameter perturbations are calculated from a basic model derived from the equilibrium and induced electron densities at a charged jellium-air interface within LRA. We have derived an analytical expression of the $d_\parallel$ component perturbation and demonstrated that it exhibits a purely classical behavior, which is equivalent to a perturbation of the surface conductivity defined with bulk parameters. Moreover, the perturbed in-plane response accounts for the observed spectral shift of the resonance frequency observed in experiments. However, the intriguing spectral narrowing of the resonance as more electrons are added to the resonator cannot be explained by our basic model. We attribute these discrepancy to unaccounted non-classical effects such as: nonlocality and anisotropy of the local permittivity. To provide stronger evidence for these claims, further modeling efforts that go beyond jellium considerations are required. Moreover, as discussed in the text, exploring optimized geometries with stronger field localization will aid to more effectively distinguish the contribution arising from the $d_\mathrm{\perp}$ component. Additionally, we observed that smaller resonators exhibit more pronounced alterations in resonance behavior, because of their increased surface-to-volume ratio. If these resonators are further downsized, this effect is anticipated to become even more prominent, offering potential for the development of electrically-driven ultra-fast optical modulators by gaining control over non-classical surface effects.

\vspace{7.5mm}
\noindent		
\textbf{Methods}

\vspace{5mm}
\noindent		
\textbf{Numerical simulations.}  The perturbation of the optical response of a plasmonic nanoresonator under electrostatic biasing is numerically determined using the commercially available FEM solver (COMSOL Multiphysics 6.0)\,\cite{hadoop} for the electrodynamics. Our simulations involve a two-step process. Initially, we solve for the electrostatic field by applying a potential $V$ to the structure using the AC/DC Module. This step allows us to obtain the induced surface electron density $\Delta\eta_\text{0}(\textbf{r}_\text{s},V)$. The ground potential is placed infinitely far away by employing an infinite element domain layer. In the second step, we conduct optical simulations using the wave optics module to analyze the scattering cross Section based on the local perturbation in the $d$-parameters influenced by the induced surface electron density. To introduce perturbations in the $d$-parameters, we employ mesoscopic boundary conditions (see Supplementary Section 1.6) implemented with an auxiliary potential method, as described in Ref.\,\onlinecite{yang2019general}. The structure is excited with a plane wave polarized along the long axis of the nanoresonator, and the scattered light is collected at the bottom hemisphere to mimic the experimental setup. The optical permittivity of gold is taken from the experimental values for mono-crystalline
gold provided by Olmon \emph{et al.}\,\cite{olmon2012optical}, while glass is modeled using Sellmeier coefficients. For more information, please refer to Supplementary Section\,1.3.

\vspace{5mm}
\noindent
\textbf{Sample fabrication.} Mono-crystalline gold microplatelets, measuring 50\,nm in thickness, are synthesized through a wet-chemical process outlined in Ref.\,\onlinecite{krauss2018controlled}. These microplatelets are then transferred onto a glass coverslip (24\,mm $\times$ 24\,mm \#1.5 Menzel) with evaporated metal layers featuring an array of electrode pads prepared by optical lithography and electron beam physical vapor deposition (20\,nm chromium adhesion layer, 80\,nm gold layer). The microplatelets are carefully positioned over the glass window on structured microscopic electrodes, ensuring a conductive connection between the flake and the metal film. Nanoresonator fabrication is conducted as described in Ref.\,\onlinecite{meier2022controlling}.

\vspace{5mm}
\noindent		
\textbf{Optical characterization.} To capture dark-field scattering spectra of plasmonic nanoresonators, we employ an inverted optical microscope (TE2000-U, Nikon) equipped with a nanopositioning piezostage (NanoLPS200, Mad City Labs Inc.) and an oil-immersion microscope objective (PlanApochromat, 100$\times$, NA = 1.45, Nikon). As excitation source, we utilize a supercontinuum laser (SuperK FIANIUM, FIR-20, NKT) that is spectrally shaped and scanned in 10\,nm increments from 500\,nm to 900\,nm using an acousto-optic tunable filter (SuperK SELECT, NKT). The light from the laser is sent through a 300\,$\upmu$m pinhole and focused to the back focal plane of the oil-immersion microscope objective via a 500\,mm lens, providing a collimated beam at the sample. To separate the detection and excitation beam paths, a 50:50 beam splitter is employed. Light scattered by the structure above the critical angle and light reflected directly from the sample are separated using a circular beam block. Additionally, to minimize potential stray light, an iris is positioned in the intermediate image plane and adjusted until the background is completely suppressed. The signal is collected using an optical power meter (1835-C, Newport). See also Supplementary Section\,2.1-2.3.

\vspace{5 mm}
\noindent
\textbf{Electro-optical measurements.} The nanoresonators are electrically connected to the macroscopic electrode pads using thin connector lines. They are then further connected via micro-manipulators to a function generator (DS 345, Standford Research Instruments). An AC voltage is applied to the structure, and the scattered light, guided to the detector, is divided into two collecting channels using a 90:10 beam splitter. The majority of the light is captured by a silicon photodetector (DET36A2, Thorlabs) with a rise time of 14\,ns, while the remaining portion is detected using an optical power meter (1835-C, Newport). The electrical signal from the photodetector is directed to a lock-in amplifier (DSP 7260, EG$\&$G), with the reference signal obtained directly from the function generator output using a T-splitter. To enable the recording of correlated data, the entire process is monitored via a LabVIEW program. Throughout the measurements, the sample is continuously blow-dried using a laminar nitrogen stream. 

\vspace{5 mm}

\noindent 
\textbf{Acknowledgments.} 
T.\,F. acknowledges funding by the Volkswagen Foundation via an ‘Experiment!’ grant (95869), by the Marie Sk\l{}odowska-Curie Actions (MSCA) individual fellowship project PoSHGOAT (project- id 837928) and participation in CA19140 (FIT4NANO), supported by COST (European Cooperation in Science and Technology). B.\,H.\,gratefully acknowledges funding by the Deutsche Forschungsgemeinschaft (DFG, German Research Foundation) under Germany’s Excellence Strategy through the Würzburg-Dresden Cluster of Excellence on Complexity and Topology in Quantum Matter ct.qmat (EXC 2147, Project ID ST0462019) as well as through a regular project (HE5618/10-1) and a Reinhard Koselleck project (HE5618/6-1). The Volkswagen foundation is acknowledged for funding via research grant 93437-1. N.\,A.\,M. is a VILLUM Investigator supported by VILLUM FONDEN (Grant No.\,16498). The Center for Polariton-driven Light--Matter Interactions (POLIMA) is funded by the Danish National Research Foundation (Project No.\,DNRF165).


\vspace{5 mm}

\noindent 
\textbf{Author contributions}  L.\,Z. conceived the idea and designed the experiment. J.\,M. fabricated the structures and performed the SEM measurements. R.\,K. and L.\,Z. built and optimized the electro-optical setup. L.\,Z. performed the electro-optical experiments. N.\,A.\,M. provided the derivation of the $d$-parameter perturbations. C.\,W. performed the DFT calculations. L.\,Z. conducted the FEM simulations. L.\,Z. analyzed the data and drafted the manuscript with input from all the authors. B.\,H. and T.\,F. supervised the project.

\vspace{5 mm}

\noindent
\textbf{Data and code} is available upon request to L.\,Z. and T.\,F.

\vspace{5 mm}

\noindent
\textbf{Correspondence} should be addressed to L.\,Z. and T.\,F.

\newpage
\bibliographystyle{unsrt}
\bibliography{bib_charging}

\newpage

\includepdf[pages={{},{},1,{},2,{},3,{},4,{},5,{},6,{},7,{},8,{},9,{},10,{},11,{},12,{},13,{},14}]{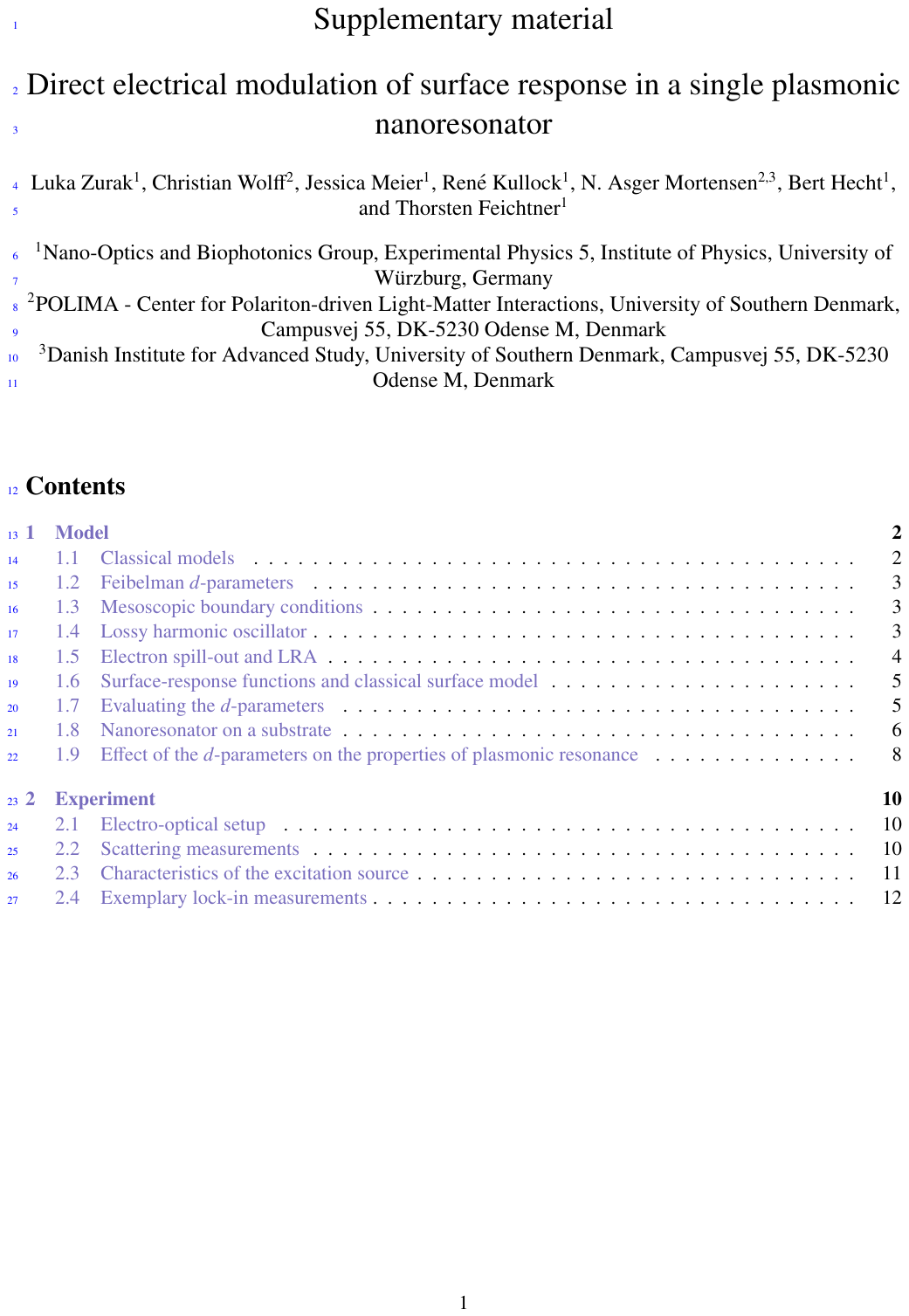}

\end{document}